\documentclass[prl,twocolumn,preprintnumbers,amsmath,amssymb]{revtex4-1}
\usepackage{commath}  
\usepackage{graphicx}
\usepackage[utf8]{inputenc}
\usepackage{xcolor}

\def\MT{\ensuremath{ m_\text{t}}}
\def\MZ{\ensuremath{ M_\text{Z}}}

\def\glu{\ensuremath{\tilde{g}}}

\newcommand{\mlog}[2]{\ensuremath{\log \frac{#1}{#2} }}

\newcommand{\ov}{\overline}
\newcommand{\Bbar}{\,\overline{\!B}}
\newcommand{\bb}{\ensuremath{B\!-\!\Bbar{}\,}}
\newcommand{\bbm}{\bb\ mixing}
\newcommand{\gev}{\ensuremath{\;\mbox{GeV}}}
\newcommand{\tev}{\ensuremath{\;\mbox{TeV}}}
\newcommand{\eq}[1]{Eq.~(\ref{#1})}
\newcommand{\eqsand}[2]{Eqs.~(\ref{#1}) and (\ref{#2})}
\newcommand{\nn}{\nonumber\\}

\begin{document}

\preprint{TTP19-016, P3H-19-02}

\title{Little hierarchies solve the little fine-tuning problem:\\ a case
  study in supersymmetry with heavy guinos}
\author{Thomas Deppisch}
\email[]{thomas.deppisch@kit.edu}
\author{Ulrich Nierste}
\email[]{ulrich.nierste@kit.edu}
\affiliation{Institut f\"ur Theoretische Teilchenphysik (TTP), Karlsruher Institut f\"ur Technologie (KIT), 76131 Karlsruhe, Germany}

\date{August 3, 2019}

\begin{abstract}
  Radiative corrections with new heavy particles coupling to Higgs
  doublets destabilize the electroweak scale and require an ad-hoc
  counterterm cancelling the large loop contribution.  If the mass scale
  $m_1$ of these new particles in in the TeV range, this feature
  constitutes the \emph{little fine-tuning problem}.  We consider the
  case that the new-physics spectrum has a little hierarchy with two
  particle mass scales $m_{1,2}$ and $m_2 ={\cal O}(10\,m_1)$ and no
  tree-level couplings of the heavier particles to Higgs doublets.  As
  a concrete example we study the (next-to-)minimal supersymmetric
  standard model ((N)MSSM) for the case that the gluino mass
  $M_3$ is significantly larger than the stop mass parameters
  $m_{L,R}$ and show that the usual one-loop fine-tuning
  analysis breaks down. If $m_{L,R}$ is defined in the
  dimensional-reduction ($\ov{\rm DR}$) or any other fundamental scheme,
  corrections enhanced by powers of
  $M_3^2/m_{L,R}^2$ occur in all higher loop
  orders. After resumming these terms we find the fine-tuning measure
  substantially improved compared to the usual analyses with
  $M_3\lesssim m_{L,R}$.
  In our hierarchical scenario the stop self-energies grow like
  $M_3^2$, so that the stop masses $m_{L,R}^{\rm OS}$
  in the on-shell (OS) scheme are naturally much larger than their
  $\ov{\rm DR}$ counterparts $m_{L,R}^{\ov{\rm DR}}$. This
  feature permits a novel solution to the little fine-tuning problem:
  $\ov{\rm DR}$ stop masses are close to the 
  electroweak scale, but radiative corrections involving the heavy
  gluino push the OS masses, which are probed in collider searches,
  above their experimental lower limits. As a byproduct, we clarify 
  which renormalization scheme must be used for squark masses
  in loop corrections to low-energy quantities such as the 
  \bbm\ amplitude. 
\end{abstract}


\maketitle

\section{Introduction}
Theoretical attempts to unify gauge forces necessarily lead to new
particles with masses way above the electroweak scale $v=174\gev$
defined by the vacuum expectation value (vev) of the Standard-Model (SM)
Higgs boson. Such heavy particles generally lead to unduly large
radiative corrections to $v^2$, in conflict with the naturalness
principle which forbids fine-tuned cancellations between loop
contribution and counterterm for any fundamental parameter in the
lagrangian \cite{Weinberg:1975gm,Susskind:1978ms,tHooft:1979rat,
  Veltman:1980mj}. The
observation that in supersymmetric field theories \cite{Wess:1974tw}
corrections to the electroweak scale vanish exactly
\cite{Kaul:1981hi,Inami:1982xb,Deshpande:1984ke} made supersymmetric
models the most popular framework for studies of beyond-Standard-Model
(BSM) phenomenology.

Supersymmetry breaking introduces a mass splitting between the SM
particles and their superpartners. Increasing lower bounds on the masses
of the latter derived from unsuccessful searches at the LEP, Tevatron,
and LHC colliders brought the fine-tuning problem back: Specifically, stops
heavier than $~1\tev$ induce loop corrections to the Higgs potential
which must be cancelled by tree-level parameters to two or more
digits. Owing to this \emph{little fine-tuning problem}\  low-energy supersymmetry
has lost some of its appeal as a candidate for BSM physics.  
Nevertheless, analyses of naturalness in supersymmetric theories,
which are under study since the pre-LEP era, still receive a lot
of attention \cite{Sakai:1981gr,
  Ellis:1986yg,Barbieri:1987fn, Chankowski:1997zh,Chankowski:1998xv,
  Barbieri:1998uv, Feng:1999zg, Kitano:2005wc, Kitano:2006gv,
  Ellis:2007by, Hall:2011aa, Strumia:2011dv,
  Baer:2012mv,Fichet:2012sn,Cabrera:2012vu, Baer:2012cf,
  Baer:2012up,Boehm:2013gst,Balazs:2013qva,Kim:2013uxa, Casas:2014eca,
  Baer:2015tva, Drees:2015aeo, Baer:2015rja, Kim:2016rsd,
  vanBeekveld:2016hug,Cici:2016oqr,Cabrera:2016wwr,Buckley:2016kvr,Baer:2017pba,
  Abdughani:2017dqs,Fundira:2017vip,Baer:2018rhs,vanBeekveld:2019tqp}.

In this paper we study the little fine-tuning problem for the case of a
hierarchical superpartner spectrum, with gluinos several times heavier
than the stops. The gluino mass is less critical for fine-tuning,
because gluinos couple to Higgs fields only at the two-loop level. In
such a scenario the usual fine-tuning analyses based on fixed order
perturbation theory break down. Denoting the left-chiral
and right-chiral stop mass parameters by 
$m_{L,R}^2$ and the gluino mass by $M_3$ we identify
$n$-loop corrections enhanced by
$\left[ M_3^2/m_{L,R}^2\right]^{n-1}$ and resum
them. 
These terms are not captured by renormalization-group (RG) analyses of
effective Lagrangians derived by successively integrating out heavy
particles at their respective mass scales, which instead target large
logarithms. 

Our findings do not depend on details of the Higgs sector, and we
exemplify our results for both the Minimal Supersymmetric Standard Model
(MSSM) and its next-to-minimal variant NMSSM. The results also trivially
generalise to non-supersymmetric theories with little hierarchies
involving a heavy scalar field coupling to Higgs fields and a heavier fermion
coupling to this scalar.

\section{Corrections to the Higgs mass parameters in the (N)MSSM}
We consider only small or moderate values of the ratio
$\tan\beta\equiv v_2/v_1$ of the vacuum expectation values (vevs) of the
two Higgs doublets $H_{1}=(h_{1}^0, h_{1}^-)^T$,
$H_{2}=(h_{2}^+, h_{2}^0)^T$, so that all Yukawa couplings are small
except for the coupling $y_t$ of the (s)tops to $H_2$. Our 
(N)MSSM loop calculations involve the gluino-stop-top vertices as well as the
couplings encoded in the superpotential
\begin{equation} \label{eq:supo}
  \mathcal{W} =  y_t\, \left( \tilde t_R \tilde t_L\, h_2^0 - \tilde t_R
    \tilde b_L \, h_2^+ \right) 
\end{equation}
and the supersymmetry-breaking Lagrangian
\begin{align}
-\mathcal{L}_{\mathrm{soft}}
  &=  A_t \, \left(\tilde t_R \tilde t_L\, h_2^0 -
     \tilde t_R \tilde b_L\, h_2^+ \right) + \mathrm{H.c.} \nonumber\\
  &\quad + m_L^2 \left(\tilde t_L^\star \tilde t_L +
    \tilde b_L^\star \tilde b_L\right) +
     m^2_{h_2} \left(h_2^{0,\star}h_2^0 + h_2^{+,\star}h_2^+\right) \nonumber\\
  &\quad + m^2_{h_1} \left(h_1^{0,\star}h_1^0 +
    h_1^{-,\star}h_1^-\right) + m_R^2\, t_R t_R^\star \nonumber \\
  &\quad+ \frac 12\, M_3\, \overline{\psi_{\glu}} \psi_{\glu} 
 \label{eq:lsoft}
\end{align}
with the stop, sbottom, and gluino fields 
$\tilde t_{L,R}$,$\tilde b_{L,R}$,$\psi_{\glu}$, respectively.
In the notation of Ref.~\cite{Ellwanger:2009dp} the
($\mathbb{Z}_3$ symmetric) NMSSM Higgs potential reads
\begin{align}
    V_{higgs} = &\abs{\kappa s^2-\lambda h_1^0 h_2^0}^2 + (m_{h_1^2}^2 +
    \lambda^2 \abs{s}^2) \abs{h_1^0}^2 \nn
    & +  (m_{h_2^2}^2 +
    \lambda^2 \abs{s}^2) \abs{h_2^0}^2  
    + \frac{g^2}{4} \left( \abs{h_2^0}^2-\abs{h_1^2}\right)^2 \nn
    & + m_s^2
    \abs{s}^2 + \left(\frac{1}{3} A_\kappa s^3 - A_\lambda h_1^0h_2^0s +
      \mathrm{H.c.}\right). \label{eq:vh}
\end{align}
Note that $g^2\equiv (g_1^2+g_2)^2/2$ and terms with charged fields
are dropped. The singlet field $s$ acquires the vev $v_s$. The
electroweak scale is represented by the $Z$ boson mass $M_Z$.
\begin{figure}[t]
\includegraphics[width=.35\linewidth]{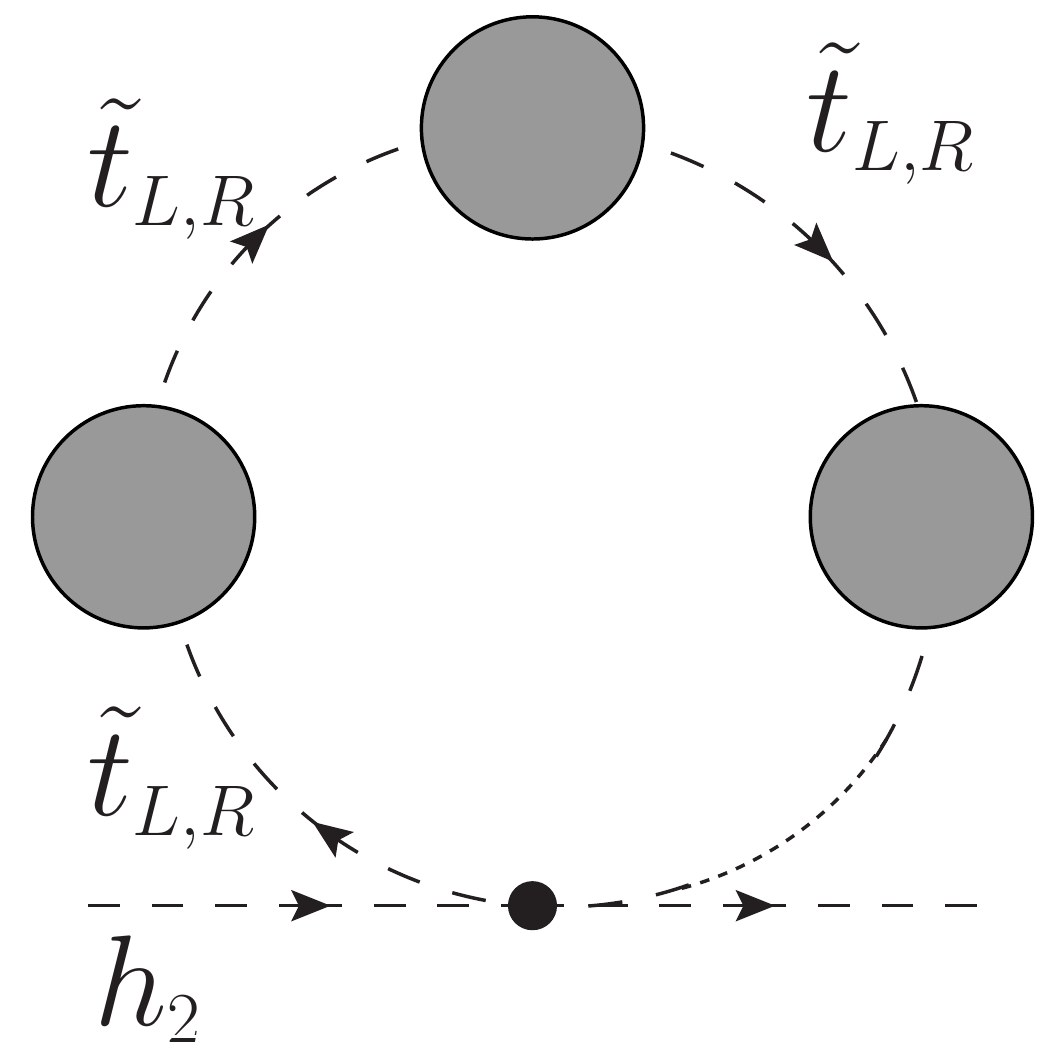}\hfill
\includegraphics[width=.55\linewidth]{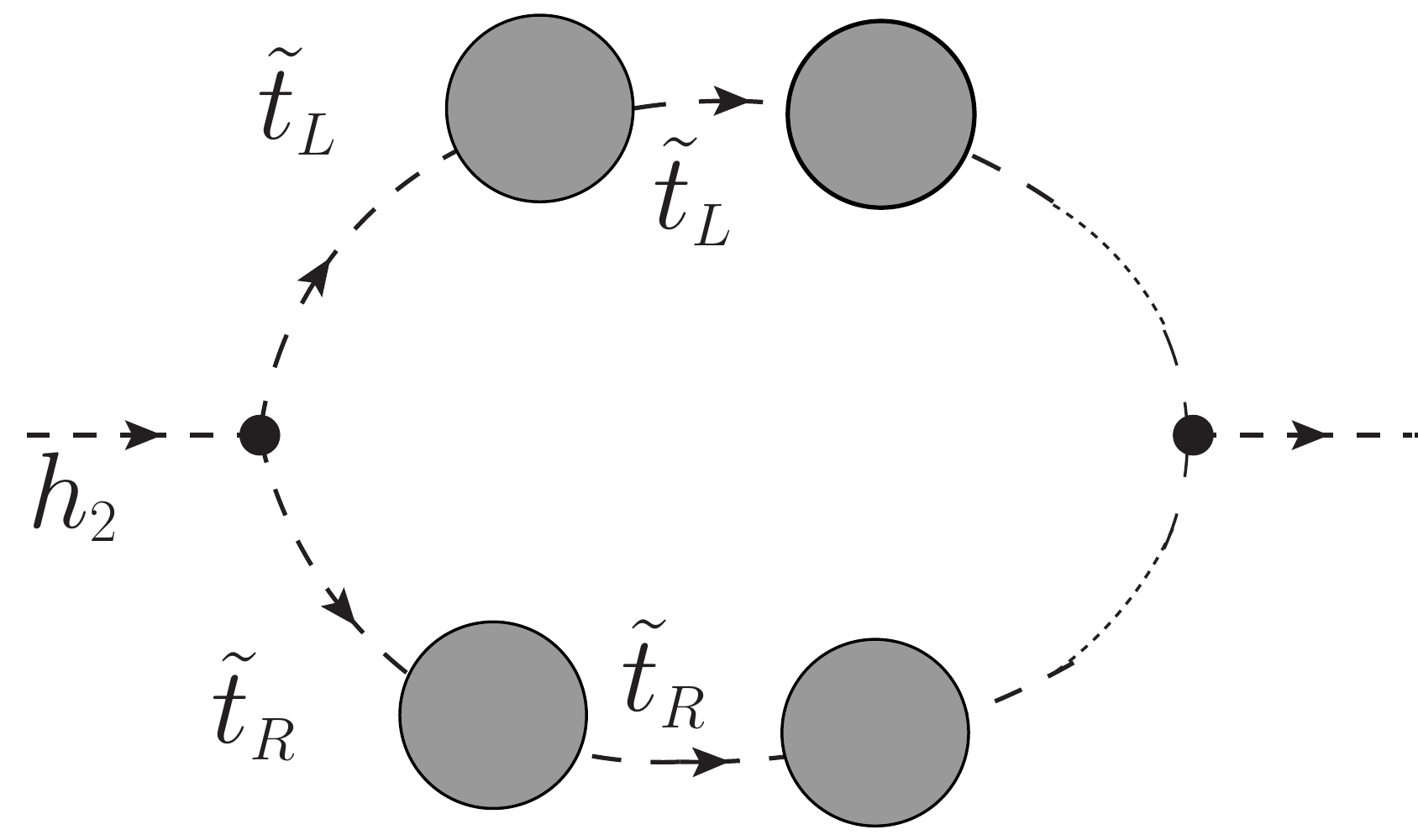}
\caption{Resummed contributions to $m_{22}^{2}$.\label{fig:sum}}
\end{figure}
Minimizing $V_{higgs}$ gives
\begin{align}
  \frac{1}{2}M_Z^2 &= \frac{ m_{11}^2 \cos^2\beta -
                     m_{22}^2 \sin^2\beta}{\sin^2\beta-\cos^2\beta} 
\label{eq:min}  
\end{align}
with the tree-level contributions
\begin{align}
   m_{11}^{2\,(0)} &= m_{h_1}^2 + \lambda^2 |v_s|^2,\quad  
              m_{22}^{2\, (0)} = m_{h_2}^2 + \lambda^2 |v_s|^2 .
  \label{eq:defm11}                       
\end{align}  
In the MSSM \eqsand{eq:vh}{eq:defm11} hold with the replacements
$\lambda s, \lambda v_s \to \mu_h $, $\lambda, \kappa,A_\kappa \to 0 $,
and $A_\lambda s, A_\lambda v_s\to B\mu_h$ with the higg\-sino mass term
$\mu_h$ and the soft supersymmetry breaking term $B\mu_h$. In the
following we identify $\mu_h\equiv \lambda v_s$ and
$B\mu_h\equiv A_\lambda v_s$, which allows us to use the same notation
for MSSM and NMSSM.  Next we integrate out the heavy sparticles and
thereby match the (N)MSSM onto an effective two-Higgs-doublet model.  We
parametrize the loop contributions as
\begin{align}
  m_{22}^2 &= m_{22}^{2\,(0)}
             + m_{22}^{2\,(1)}  + m_{22}^{2\,(2)}
             + m_{22}^{2\,(\geq 3)}
  \label{eq:m22exp}                       
\end{align} 
with the well-known one-loop term 
\begin{align}
      m_{22}^{2,(1)} = &
      -\frac{3\, \abs{y_t}^2}{16\, \pi^2} \left[ m_L^2
        \left(1-\mlog{m_L^2}{\mu^2}\right) + L\to R 
      %
                         \right]
                         \nn                      
      &\hspace{-2em} -\frac{3\, \abs{A_t}^2}{16\, \pi^2}
      \frac{m_R^2 - m_R^2\mlog{m_R^2}{\mu^2} -m_L^2 +
        m_L^2 \mlog{m_L^2}{\mu^2}}{m_L^2-m_R^2}  \label{eq:m221}
\end{align}
in the modified dimensional reduction ($\ov{\rm DR}$) scheme.
$\mu={\cal O} (m_{L,R})$ is the renormalization scale. The corrections
to other mass parameters like $m_{11}^2$ are small as long as $|A_t|,|\mu_h|$
are not too large. At one-loop order the fine-tuning issue only concerns
the first term in $m_{22}^{2,(1)}$, which requires sizable cancellations with
$m_{22}^{(0)}$ to reproduce the correct $M_Z$ in \eq{eq:min}.  

\begin{figure}[t]
\includegraphics[width=\linewidth]{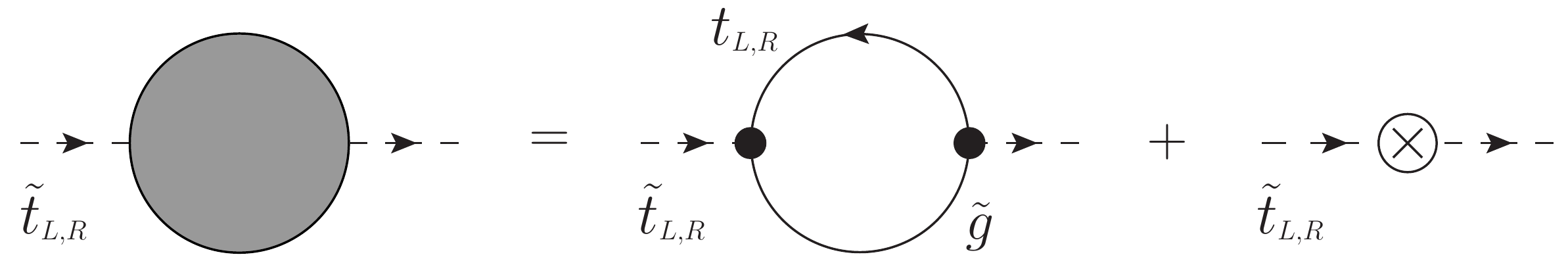}
\caption{Stop self-energies with gluino loop and counterterm.\label{fig:se}}
\end{figure}

At $n$-loop level with $n\geq 2$ we only consider the contributions
enhanced by $ \left( M_3^2/m_{L,R}^2 \right)^{n-1} $ with respect to
$ m_{22}^{2\,(1)}$ stemming solely from Feynman diagrams with $n-1$ stop
self-energies shown in Fig.~\ref{fig:sum}. Other multi-loop diagrams
involve fewer stop propagators and do not contribute to the highest
power of $M_3^2/m_{L,R}^2$.  The self-energies involve a gluino-top loop
and a stop mass counterterm, see Fig.~\ref{fig:se}.
  We decompose $ m_{22}^{2\,(n)}$ as
  \begin{align}
      m_{22}^{2\,(n)} &=       m_{22\, I}^{2\,(n)} \, +\,       m_{22\, II}^{2\,(n)} 
  \end{align}
  for the two sets of diagrams in  Fig.~\ref{fig:sum}. The left diagrams
  constituting $ m_{22\,
    I}^{2\,(n)}$ have $n$ stop propagators while the right ones summing
    to $ m_{22\, II}^{2\,(n)}$ have
    $n+1$ stop propagators.
  Inspecting the UV behaviour of the stop loop shows that 
  only  $ m_{22\,
    I}^{2\,(2)}$ contains a 
  logarithm
  $\log (M_3/m_{L,R})$.  Explicit calculation of the two-loop diagrams yields 
\begin{align}
  m_{22}^{2\,(2)} =&
  \frac{\alpha_s(\mu) \,\abs{y_t}^2 
    M_3^2}{4\,\pi^3}\biggl[   \nn 
& \quad 
      -  \left(1+  \mlog{\mu^2}{M_3^2} \right)
       \left(1+2\mlog{\mu\,M_3}{m_L\, m_R} \right) \nn
& \qquad                                                    \qquad                                                    
  + \frac{\pi^2}{3} +{\cal O} \left( \frac{m_{L,R}^2}{M_3^2}\right) \biggr]
\label{eq:m222}
\end{align}
in the ($\ov{\rm DR}$) scheme. 
If one considers very large mass splitting between $m_{L,R} $ and $M_3$,
one may choose to integrate out these sparticles at different scales and
finds $\mu\sim M_3$ more appropriate than $\mu={\cal O} (m_{L,R})$ in
$\alpha_s(\mu)$ and the first logarithm in \eq{eq:m222}.
$ m_{22\, II}^{2\,(2)}$ has no $\log (M_3/m_{L,R})$ and amounts to only
$\sim 10\%$ of $ m_{22\, I}^{2\,(2)}$ for the numerical examples
considered below.

For $M_3\gg m_{L,R}$ we find for the resummed higher-order contributions:
\begin{align}
  m_{22\, I}^{2\,(\geq 3)} &= \frac{3\,\abs{y_t}^2 }{16\pi^2} m_L^2
                      \sum_{k=2}^\infty  \frac{\xi_L^k}{k(k-1)} + L\to
                         R \nn
  & \!\!\!\!\!\!\!
    = \frac{3\,\abs{y_t}^2 }{16\pi^2} m_L^2 \left[\xi_L +(1-\xi_L)
      \log(1-\xi_L) \right]  + L\to
      R \label{eq:m22n1} \\
  m_{22\, II}^{2\,(2)} &+  m_{22\, II}^{2\,(\geq 3)} 
                      \, = -\frac{3\, \abs{A_t}^2}{16\pi^2}
                              \sum_{k=1}^\infty \frac{\xi_{L,R}^k}{k} 
                         \nn
    &\qquad\qquad\;\,
      = \frac{3\, \abs{A_t}^2}{16\pi^2} \log(1-\xi_{L,R})  
      \label{eq:m22n2} 
\end{align}
with
\begin{align}
 \xi_{L,R} &\equiv& - \frac{4 \alpha_s(\mu)}{3\pi} \frac{M_3^2}{m_{L,R}^2} 
             \left[ 1+ \mlog{\mu^2}{M_3^2}\right] \, + \, \Delta \xi_{L,R}.
 \label{eq:xi}                   
\end{align}
$ \Delta \xi_{L,R}$ controls the renormalization scheme of the stop
masses, $ \Delta \xi_{L,R}=0$ for the $\ov{\rm DR}$ scheme.  For
simplicity we quote the numerically less important term in \eq{eq:m22n2}
for the special case $m_L=m_R$.  For $M_3\sim 5\, m_{L,R}$ one finds
$\xi_{L,R}\sim -1$, so that $ m_{22\, I,II}^{2\,(\geq 3)}$ is of similar
size as $ m_{22\, I,II}^{2\,(1)}$.  The expressions above define
$m_{22, I,II}^{2\,(n)}$ at the scale $\mu\sim m_{L,R}$. We minimize the
Higgs potential at the lower scale $m_t$ (denoting the top mass) where
\begin{align}
m_{22}^2 (\MT) &=  
 \Bigl(1 - \frac{6\, \abs{y_t}^2}{16\,\pi^2} \mlog{\mu}{\MT} \Bigr)\, m_{22}^2\;(\mu) \;,
\end{align}
while the running of $m_{11}^2$ and $m_{12}^2\equiv B\mu_h$ is
negligible.  

Next we switch to the on-shell (OS) scheme for the stop masses. For
clarity we consider the case of small $|A_t|$ and $|\mu_h|$, so that stop
mixing is negligible and $m_{L,R}^{\rm OS}$ coincide with the two mass
eigenstates. In the OS scheme the counterterm $ \Delta \xi_{L,R}$ in
\eq{eq:xi} cancels the stop self-energies and renders $\xi_{L,R}=0$. 
Thus $ m_{22}^{2\,(\geq 3)}=m_{22,II}^{2\,(2)}=0$, while
$m_{22,I}^{2\,(2)}$ is non-zero due to the different UV behavior of the
stop momentum loop:
\begin{align}
  m_{22}^{2,(2)\,\rm OS} &=
          \frac{\alpha_s(\mu) \,\abs{y_t}^2 M_3^2}{4\,\pi^3}
             \biggl[
                    -1 + \mlog{\mu^2}{m_L\,m_R} \nn
& \qquad                    + \log^2 \frac{\mu^2}{M_3^2} + \frac{\pi^2}{3}
          + {\cal O} \left( \frac{m_{L,R}^2}{M_3^2}\right) 
                 \biggr] 
                           \label{eq:m222os}
\end{align}
Thus with stop pole masses no $M_3^2/m_{L,R}^2$ enhanced terms appear
beyond two loops and the resummation of the higher-order terms is
implicitly contained in the shift $m_{L,R} \to m_{L,R}^{\rm OS}$, which
absorbs the higher-order terms into $m_{22}^{(1)}$ and
$m_{22}^{(2)}$. The $\mu$ dependence in \eq{eq:m222os} results from the
stop loop integration, i.e.\ the superscript ``OS'' in \eq{eq:m222os}
only refers to the definition of the stop mass, while $ m_{22}^{2}$ is
still $\ov{\rm DR}$ renormalized.

For the fine-tuning issue there are several important lessons: Most
importantly, $m_{L,R}^{2\, \rm OS}$ is \emph{larger} than $m_{L,R}^2$ by
terms $\propto \alpha_s M_3^2$, meaning that the LHC lower bound on
$m_{L,R}^{\rm OS}$ permits a $\ov{\rm DR}$ mass $m_{L,R}$ closer to the
electroweak scale complying with naturalness. That is,
$m_{L,R}^{\rm OS}$ could well be dominated by the gluino-top
self-energy. In the on-shell scheme we observe moderate fine-tuning in
$m_{22}^2$ if we vary $m_{L,R}$, partly because the large radiative
piece of $m_{L,R}^{\rm OS}$ depends only logarithmically on $m_{L,R}$,
and partly because the effects from $m_{22}^{2\, (1)}$ and
$m_{22}^{2\, (2)}$ have opposite signs and tend to cancel out. This
behavior can be better understood if we solely work in the $\ov{\rm DR}$
scheme: For $m_{L,R}$ close to the electroweak scale none of the
infinite number of terms $m_{22}^{(n)}$ is individually so large that it
calls for a fine-tuned $m_{22}^{(0)}$ in \eq{eq:m22exp}. We may instead
be concerned about the fine-tuning related to a variation of $M_3$: In a
perturbation series truncated at order $n$ we see a powerlike growth
with terms up to $\xi_{L,R}^n$ in the sum in \eq{eq:m22n1}, with the
terms of different loop orders having similar magnitude and alternating
signs.  However, the resummation tempers this behaviour to
$m_{L,R}^2 \xi_{L,R} \sim M_3^2$. We have numerically checked that we
obtain the same results for $m_{22}^{2}$ in both approaches, i.e.\
by either employing the explicit resummation in the  $\ov{\rm DR}$
scheme or converting the stop masses to the OS scheme. 

\section{Numerical study of the fine-tuning}
We use the Ellis-Barbieri-Giudice 
\begin{figure}[t]
  \includegraphics[width=.95\linewidth]{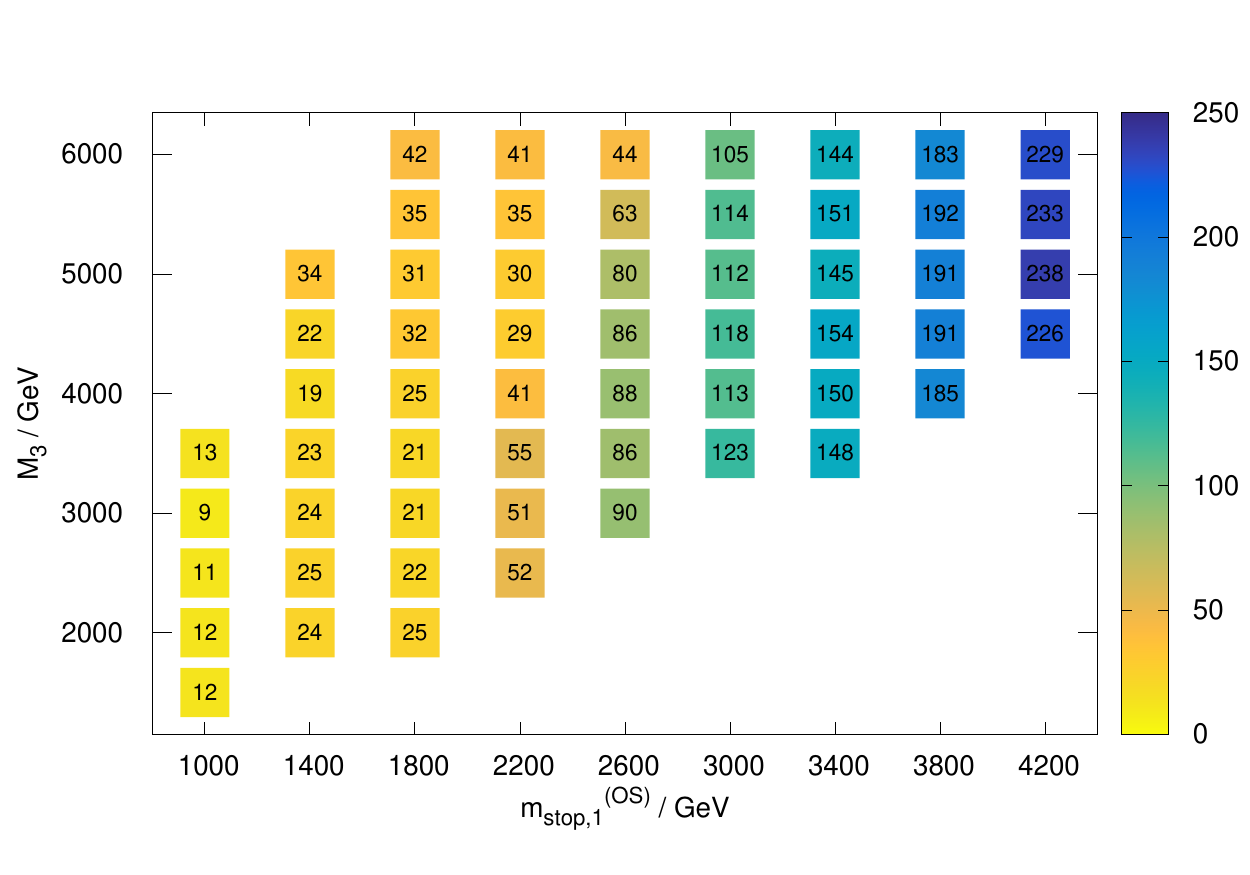}
\caption{Fine-tuning measure $\Delta(m_L)$ for different values 
 of the lighter on-shell stop mass (essentially equal to $m_L^{\rm OS}$ in our 
 analysis) and $M_3$. The number gives the mean of 100 sample points that
 correctly reproduce $M_Z=91\gev$ and $m_h=125\gev$
 \cite{Chatrchyan:2012xdj,Aad:2012tfa}. \label{fig:nmssm}
}
\end{figure}
fine-tuning measure \cite{Ellis:1986yg,Barbieri:1987fn} 
\begin{equation} \label{eq:FT}
  \Delta(p) = \abs{ \frac{p}{\MZ(p)}\, \frac{\partial\, \MZ(p)}{\partial p}} \;,
\end{equation}
where $p$ stands for any Lagrangian parameter.
Using $\ov{\rm DR}$ stop masses as input we calculate the $\ov{\rm OS}$
masses which enter the loop-corrected Higgs potential through
\eqsand{eq:m221}{eq:m222os}. For the latter we determine all two-loop
contributions to $m_{11}^2$, $m_{12}^2$, and $m_{22}^2$ involving
$\alpha_s$, $y_t$, $A_t$ exactly. E.g.\ we go beyond the large-$M_3$
limit of the previous section and calculate 205 two-loop diagrams in
total.  For this we have used the Mathematica packages \texttt{FeynArts}
\cite{Hahn:2000kx} (with the Feynman rules of Ref.~\cite{Rosiek:1995kg})
and \texttt{Medusa} \cite{cwiegand1,cwiegand2}, which performs
asymptotic expansions in small external momenta and large masses. The
analytic methods involved are based on
Refs.~\cite{Davydychev:1992mt,Nierste1993, Fleischer:1994ef,
  Davydychev:1995nq, Nierste:1995fr,Anastasiou:2006hc}.

We start with the discussion of the NMSSM: With two of the three
minimization conditions we trade the parameters $m_s^2$ and $A_\lambda$ for
$\mu_h\equiv\lambda v_s$ and $\tan\beta$. (The third minimization condition is
\eq{eq:min} yielding $M_Z$.) For the illustrative example in
Fig.~\ref{fig:nmssm} we fix the parameters $\tan\beta=3$,
$\lambda=0.64$, $\kappa=0.25$, $\mu_h=200\, \mathrm{GeV}$, and
$m_{11}^{(0)} =600\, \mathrm{GeV}$. Then we choose $m_{22}^{2,(0)}$,
$A_t$, $m_L^{\ov{\rm DR}}$, $m_R^{\ov{\rm DR}}$, $A_\kappa$ randomly
subject to the contraints that the correct values of $M_Z$ and the
lightest Higgs mass $m_h=125\gev$ as well as the smaller stop mass
$m_{\tilde t,1}^{\rm OS}$ displayed in Fig.~\ref{fig:nmssm} are
reproduced for a given value of $M_3$. We calculate $\Delta(m_L)$ for over 100
different parameter points corresponding to a given point
$(m_{\tilde t,1}^{\rm OS},M_3)$; the number in the colored square is
the average $\Delta(m_L)$ found for these points. For most of our parameter points 
$m_{\tilde t,1}^{\rm OS} \approx m_L$, but this feature is irrelevant   
because the formulae are symmetric under $m_L \leftrightarrow m_R$. 
By quoting the average rather than the minimum of $\Delta(m_L)$ 
we make sure that a good fine-tuning measure is not due to accidental
cancellations. 

To illustrate the result of Fig.~\ref{fig:nmssm} with an example we consider
the parameter point with 
\begin{align}
  m_{11}^{(0)} &= 600\, \mathrm{GeV} & m_{22}^{(0)} &= 94\, \mathrm{GeV}
  &  M_3 & = 3\tev \nn
  A_\kappa &= -6.5\, \mathrm{GeV} &      
  A_t &= 453\, \mathrm{GeV} && \nn
  m_L^{\overline{\mathrm{DR}}} &= 611\, \mathrm{GeV}
  & m_R^{\overline{\mathrm{DR}}} &= 902\, \mathrm{GeV} \label{eq:bp}
\end{align}
which yields $m_{\tilde t,1}^{\rm OS} = 1\tev$, lying substantially
above $m_L$. Note that $M_3/m_L\approx 5$, while the hierarchy in the
physical masses is moderate, $M_3/m_{\tilde t_1}^{\rm OS} = 3$.
The fine-tuning measures for this benchmark point are   
$\Delta(m_L)=6.0$,  $\Delta(m_R)=10.8$, $\Delta(M_3)=6.3$,
$\Delta(A_t)=0.2$, and all other $\Delta(p)$ are negligibly small. 

Next we briefly discuss the MSSM. A recent analysis has found
values of $\Delta \equiv \max_p \Delta (p) \geq 63$ for  
special versions of the MSSM in scans over the parameter spaces
\cite{vanBeekveld:2019tqp}. Compared to the NMSSM
one needs larger stop masses to accomodate $m_h=125\gev$, which
then leads to larger values of  $\Delta$. Yet also for the MSSM
the hierarchy $M_3 \gg m_{L,R}$ with proper resummation of higher-order
terms improves $\Delta$. We exemplify this with the parameter point
\begin{align}
  m_{11}^{(0)} &= 1583\,\gev &  m_{22}^{(0)} &= 124\,\gev \nn
  \mu_h &= 400\,\gev & \tan\beta &= 5\nn
    M_3 &= 4500\,\gev & A_t &= 3370\,\gev \nn
  m_L &= 2787\,\gev &  m_R &=1435\,\gev\nonumber
\end{align}
The on-shell stop masses for this point are
$m_{\tilde t_1}^{\rm OS}= 2168\gev$ and
$m_{\tilde t_1}^{\rm OS}= 3012\gev$.  Despite these large masses the
fine-tuning measures $\Delta(m_L) = 13$, $\Delta(m_R)=25$,
$\Delta(M_3)=8$ have moderate values while a fine-tuning measure
$\Delta(A_t)=41$ reflects the large $A_t$ needed to accomodate
$m_h=125\,\gev$. 

Finally we remark that also low-energy observables like the \bbm\
amplitude or the branching ratios of rare meson decays (such as
$b\to s \gamma$, $K\to \pi \nu \ov{\nu}$) involve higher-order
corrections enhanced by a relative factor of $M_3^2/m_{L,R}^2$, if the
stop masses are renormalised in a mass-independent scheme like
$\ov{\rm DR}$.  This remark applies to supersymmetric theories with
minimal flavor violation (MFV) in which the leading contribution 
is dominated by a chargino-stop loop and the gluino is relevant only
at next-to-leading order and beyond.  The resummation of the gluino-stop
self-energies on the internal stop lines is trivially achieved by
using the on-shell stop masses in the leading-order prediction, because
the flavor-changing loop is UV-finite; i.e.\ we face the same situation
as with $m_{22}^{2\,(\geq 3)}$. Thus low-energy observables
effectively probe the same stop masses as the collider searches at high
$p_T$.

\section{Conclusions}
We have investigated the fine-tuning of the electroweak scale in models
of new physics with a heavy and hierarchical mass spectrum.  Studying
supersymmetric models with $M_Z < m_{L,R} < M_3$ we have demonstrated
that the usual fine-tuning analysis employing fixed-order perturbation
theory breaks down for $M_3 \sim 5\, m_{L,R}$. Resumming terms enhanced
by $M_3^2/m_{L,R}^2$ tempers the fine-tuning. This behavior is
transparent if the stop masses are renormalized on-shell: The
resummation is then encoded in the shift from the $\ov{\rm DR}$ masses
to the larger on-shell masses and new allowed parameter ranges with
small values of $m_{L,R}^2$ emerge, because large radiative corrections
proportional to $\alpha_s M_3^2$ push the physical on-shell masses over
the experimental lower bounds. In these scenarios the heavy stops are
\emph{natural}, as their masses are larger than the --parametrically
large-- self-energies. As a byproduct we have found that low-energy
observables probe the on-shell stop masses.

\begin{acknowledgments}
\paragraph{Acknowledgements.}
We thank Stefan de Boer for checking expressions
\eqsand{eq:m22n1}{eq:m22n2} and several helpful discussions and
acknowledge the support of \emph{Deutsche Forschungsgemeinschaft}\
(DFG, German Research Foundation) through RTG 1694 and grant 396021762 -
TRR 257 ``Particle Physics Phenomenology after the Higgs Discovery''.
\end{acknowledgments}

\bibliography{literature}

\end{document}